\documentclass[12pt,preprint]{aastex}
\usepackage{epsfig}
\usepackage{lscape}

\makeatletter

\newcommand{\Rmnum}[1]{\expandafter\@slowromancap\romannumeral #1@}
\makeatother

\def\Lsun{\hbox{\it L$_\odot$}}

\def\Msun{\hbox{\it M$_\odot$}}

\newcommand{\etal}{\mbox{et al.}}

\newcommand{\program}[1]{{\tt {#1}}}
\def\program{\texttt}

\begin{document}
\title{12 New Galactic Wolf-Rayet Stars Identified via 2MASS$+$\textit{Spitzer}/GLIMPSE}

\shorttitle{}

\author{Jon C. Mauerhan\altaffilmark{1}, Schuyler D. Van Dyk\altaffilmark{1}, Pat  W. Morris\altaffilmark{2}}

\altaffiltext{1}{Spitzer Science Center, IPAC, California Institute of Technology, M/C 220-6, Pasadena, CA 91125, USA}
\altaffiltext{2}{NASA Herschel Science Center, IPAC, California Institute of Technology, M/C 100-22, Pasadena, CA 91125, USA}
\begin{abstract}
We report new results from our effort to identify obscured Wolf-Rayet stars in the Galaxy. Candidates were selected by their near-infrared (Two Micron All Sky Survey; 2MASS) 
and mid-infrared (\textit{Spitzer}/GLIMPSE) color excesses, which are consistent with free-free emission from ionized stellar winds and thermal excess from hot dust.
 We have confirmed 12 new Wolf-Rayet stars in the Galactic disk, including 9 of the nitrogen subtype (WN), and 3 of the carbon subtype (WC); this raises the total number of Wolf-Rayet stars 
 discovered with our approach to 27.  We classify one of the new stars as a possible dust-producing WC9d+OB{\sc I} colliding-wind binary, as evidenced by an infrared excess resembling that of known WC9d stars, the detection of OB{\sc I} features superimposed on the WC9 spectrum, and hard X-ray emission detected by \textit{XMM-Newton}. A WC8 star in our sample appears to be a member of the stellar cluster Danks 1, in contrast to the rest of the confirmed Wolf-Rayet stars that generally do not appear to reside within dense stellar clusters. Either the majority of the stars are runaways from clusters, or they formed in relative isolation. We briefly discuss prospects for the expansion and improvement of the search for Wolf-Rayet stars throughout the Milky Way Galaxy.

 \end{abstract}
\section{Introduction}
Wolf-Rayet stars (WRs) are the post main-sequence descendants of the most massive stars ($M_{ini}\gtrsim25 \Msun$). 
They are hot ($T_{eff}$=30,000--140,000 K; Hamann et al. 2006), luminous ($L_{bol}\sim10^5$--$10^6 \Lsun$), and owing to their evolved state,
exhibit the products of core CNO processing at their surfaces. For the duration of the WR phase (few$\times$10$^{5}$ yr), dense winds are radiatively accelerated to hypersonic velocities, 
in the range $v\approx800$--$3500$ km s$^{-1}$, resulting in high mass-loss rates of $\dot{M}\sim10^{-5}$--$10^{-2}$ \Msun~yr$^{-1}$ (e.~g., see Crowther 2008; Hillier 2008). 
Thus, as sources of ionization and mechanical energy,  WRs have a profound influence on the surrounding ISM over many pc. 

WRs are characterized by strong emission lines in their spectra, which are generated and velocity-broadened 
within their fast, extended winds. They are classified into subtypes of WN (nitrogen type), WC (carbon type) and WO (oxygen type), 
based upon the relative dominance of N, C, and O emission-lines in their spectra. As WRs evolve and shed material, they become increasingly H and He deficient, and expose 
deeper and hotter regions of their atmospheres, and so are thought to follow an evolutionary sequence of 
WN$\rightarrow$WC$\rightarrow$WO (e.g., Langer et al.~1994; Norci et al.~1998). 

In our Galaxy, optical emission-line (He {\sc{ii}} $\lambda$4686) surveys have identified nearly all WR stars with $B \lesssim 14$ 
mag (Shara et al.~1999). However, the number of WRs identified in the optical ($\approx200$; van der Hucht 2003) constitute a small fraction
of the total predicted WR population, the majority of which is expected to be optically obscured by dust in the Galactic plane. Fortunately, our ability to detect WRs is greatly enhanced with infrared observations, which effectively penetrate the ISM (e.g., at 2 $\mu$m, $A_K/A_V \approx10$\%). Shara et al.~(1999) predict that 97\% of Galactic WR stars should have $K<18$ mag and, depending on the adopted WR distribution model, $\approx$50\% should have $K\lesssim14$ mag. Thus, the majority of WRs should be accessible with existing instrumentation, while roughly half of them should have be present in the 2MASS catalog of Cutri et al. (2008). Most searches for obscured WRs have been executed using near-infrared, narrow-band imaging searches, which isolate strong emission lines (Blum \& Damineli 1999; Homeier et al.~2003a, b). Although successful, such observations are expensive time-wise and limited in coverage. Fortunately, the near-infrared and mid-infrared continua of WRs provide an alternative means of selection, allowing WR searches to take advantage of recent near- and mid-infrared surveys of the Galactic plane, e.~g., by 2MASS and \textit{Spitzer}/GLIMPSE (Benjamin et al. 2003). 

All WR stars exhibit infrared excess. Figure 1 demonstrates that WRs are clearly separable from the vast majority of field stars 
in $K_s-[8.0~\micron]$ versus $J-K_s$, and $[3.6]-[8.0]$ versus $[3.6]-[4.5]~{\micron}$ color spaces. The locus of WRs within these color spaces form the basis
of a selection criterion for isolating new candidate WRs within the area of sky covered by the 2MASS$+$\textit{Spitzer}/GLIMPSE surveys. Hadfield et al. (2007) used these color criteria, only selecting sources with the highest quality 2MASS photometry (a ``AAA" photometric-quality flag of sorts), to define a sample of $\approx$5000 candidate WRs detected in the GLIMPSE-I survey, which covered an area of the Galactic plane bound by  $l=10\arcdeg$--$65\arcdeg$ and $295\arcdeg$--$350\arcdeg$, with $b= \pm 1\arcdeg$; follow-up spectroscopy resulted in the detection of 15 new WRs and $\approx$120 other hot, evolved massive stars. 

The continuum color excess in WRs is mainly due to free-free emission from e$^-$ scattering off of He$^{+}$ ions in the outer regions of their dense  winds (Wright \& Barlow 1975; Cohen et al. 1975; Morris et al. 1993).  In general, the infrared free-free excess of WRs becomes increasingly important with wavelength; the free-free spectrum rises as the wavelength increases from the near- to mid-infrared, while the Rayleigh-Jeans stellar continuum decreases. The free-free continuum excess is also supplemented by contributions from strong, broad emission lines that can affect the broad band fluxes of WRs, and produce scatter in the overall color trend. However, a significant emission-line--induced reddening or blueing of the colors in Figure 1, relative to the overall color trend, would require that emission lines have a relatively high flux in one particular band.  

Thermal emission may also significantly alter the broad-band fluxes, as in the case of some WC stars that are associated with hot dust (Cohen et al. 1975). The outlier in Figure 1 (left panel) with very red $K_s-[8.0]$ color is the dusty WC7d+O9{\sc III} binary WR 125, and so has a very large excess at 8 {\micron}. This implies that the 2MASS$+$\textit{Spitzer}/GLIMPSE WR selection criteria, as it currently stands, may be biased against WRs with very large thermal excesses due to dust. However, we are not biased to the exclusion of all WRs with thermal dust emission, since the datapoint lying closest to the upper-right area of Figure 1 (left panel), within the WR selection area, represents a new WR that is associated with thermal dust excess  (see \S5.2). 

In a forthcoming paper, we will present the results of a more detailed investigation of the various phenomenology contributing to the near- to mid-infrared excesses in Galactic WRs and other evolved massive stars, including emission lines and thermal dust emission. As a result, we will also continue to refine the color selection criteria for discovering new WRs. This work is simply a continuation of the spectroscopic follow-up of WR candidates identified using the current selection algorithm. We report on the discovery of 12 additional WRs.  

\section{Observations}
Spectra of candidate WRs were obtained in the optical and near-infrared at several facilities over the past 5 years. 
 Table 1 lists each confirmed WR star, the observation dates, and the associated facilities and instruments. Table 2 lists the $R$-band optical (USNO B1.0), near-infrared (2MASS), and mid-infrared (\textit{Spitzer}/IRAC, in most cases) photometry of each WR star.  In all of the observing runs listed in Table 1, we obtained spectra of a total of 189 candidate stars. Of these, 77 candidates were selected via the 2MASS+\textit{Spitzer}/GLIMPSE criteria, and 112 candidates were observed prior to the completion of \textit{Spitzer}/GLIMPSE, in which our selection criteria was based only on the combination of 2MASS and 8 {\micron} photometry from the \textit{Midcourse Space Experiment} (MSX; Price et al. 2001).  Considering only sources selected via the 2MASS+\textit{Spitzer}/GLIMPSE criteria, we have detected 10 WRs out of 77 observed candidates. The combination of these results with those of Hadfield et al. (2007), who report 15 WRs out of 184 candidates using similar selection criteria, implies a WR detection rate of 10\%.

As in Hadfield et al. (2007), the majority of optical and near-infrared, non-WR detections exhibited strong H$\alpha$ and Br$\gamma$ emission, respectively.  These H {\sc i} emission features are characteristics of massive OB supergiants and hypergiants. We refer the reader to Hadfield et al. (2007, their Figure 7) for an examples of the spectra of these stars. The proximity of these stars to WRs in color space, and comparison of their stellar parameters with those of WRs, will be presented by Morris et al. (2009, in prep).  Here, we focus solely on the WR confirmations.

\subsection{Optical Spectroscopy}
Optical spectra of 31 candidates were obtained at CTIO Blanco 4 m telescope on 2008, July 23--25, 2008, using the RCSpec spectrograph. The data were obtained using the KPGL3 grating with
a 1\arcsec\ slit, which provided a dispersion of $\sim$1.2\,~\AA/pixel. The spectrum covered a wavelength range of $\sim$4000--7500~\AA. Out  of 31 candidates, 2 broad emission-line stars were detected.
Spectra of 112 candidates selected from 2MASS+MSX were also obtained with the Hale 5 m telescope at Palomar on 2003, March 6--7 using the Double Spectrograph. Grating 600 was used, providing spectral coverage from 4000--8100~\AA, and dispersion of $\approx10$ ~\AA/pixel. Only one broad emission-line star was detected during this run.   
 
\subsection{Near-Infrared Spectroscopy}
Spectra of 17 WR candidates were obtained using the 3.9 m Anglo-Australian Telescope (AAT) on Siding Spring Mountain (Mount Woorat) in New South Wales, Australia, between 2008, July 14--17.  The IRIS2 instrument (Tinney et al. 2004) was used, providing a resolution of $R\approx2400$ in the $K$-band. The telescope was nodded in an ``ABBA" sequence for sky subtraction. Quartz and Xe arc lamp spectra were obtained for flat fields and wavelength calibration, prior to observing. A telluric calibration spectrum was obtained from observations of the A0 {\sc V} star HD 155379. Basic data reduction and spectral extraction were executed by the \program{Starlink ORACDR} pipeline. Telluric correction was applied to the data using the IDL program \program{xtellcor\_general} (Vacca et al. 2003). Out of the 17 stars observed, 5 exhibit broad emission lines typical of WRs.

Spectra of 29 candidates were obtained at the Southern Observatory for Astrophysical Research (SOAR), located on Cerro Pachon in Chile;  23 sources were observed in 2006 and 6 sources were observed in 2008. The Ohio State Infrared Imager Spectrometer (OSIRIS; Depoy et al. 1993) was used in cross-dispersed mode, which provided $R$$\approx$1200  in the $K$-band. Stellar spectra were acquired in an ``ABBA" nodding sequence for sky subtraction. For the 2006 run, the A0 {\sc V} star HD 106308 was observed for telluric calibration. Reduction, calibration and spectral extraction were performed using standard \program{IRAF} routines. Telluric division was performed using \program{xtellcor\_general}. Very poor weather during the 2008 run prevented us from observing a standard star, subsequent to our successful detection of 3 WRs (see \S3.2). 

\section{Spectral Classification}

\subsection{Optical Identifications}
We classify the optical spectra using well-defined emission-line diagnostics (Smith et al. 1996; Crowther, De Marco, \& Barlow 1998). Table 3 lists the equivalent widths (EW) and FWHM of prominent emission-line features in the optical stellar spectra. EW measurements were obtained using standard \program{IRAF} routines. Uncertainties are $\approx$10\% of the EW values and are the the result of imperfect fitting of the spectral lines and noise. 

Figure 2 displays the optical spectra of 2MASS J11555211$-$6245022 and 2MASS J16441069$-$4524246. The spectra of both stars exhibit lines of He {\sc i}, He {\sc ii}, N {\sc iii}, C {\sc iii}, C {\sc iv}, and a blend of H$\alpha$ and He {\sc ii}.  He and N produce the most prominent lines in the spectra of these stars, typical of WN stars. The WN subtypes can be estimated using the ratio of EWs for He {\sc ii} ($\lambda$5411~\AA)/He {\sc i} ($\lambda$5875~\AA), C {\sc iv} ($\lambda$5808~\AA)/He {\sc ii} ($\lambda$5411~\AA), and C {\sc iv} ($\lambda$5808~\AA)/ He {\sc i} ($\lambda$5875~\AA) (Smith, Shara \& Moffat 1996, see their Table 4a). For 2MASS J11555211$-$6245022, the only subtype that is consistent with all of the above emission-line ratios listed in Table 3 is WN6, so we classify this star as such. For 2MASS J16441069$-$4524246, the line ratios have significantly smaller values, and are consistent with WN7--8 subtype.

The spectra of 2MASS J13125770$-$6240599 and 2MASS J18281180$-$1025424 are displayed in Figure 3 (\textit{left panel}). These stars exhibit strong emission lines of C {\sc iii} and C {\sc iv}, typical features of WC stars. For 2MASS J13125770$-$6240599, there is also O {\sc v} emission detected at $\lambda$5590~\AA, which is useful as a WC subtype diagnostic, when compared with C {\sc iii} ($\lambda$5696 ~\AA) and C {\sc iv} ($\lambda$5801~\AA) (Crowther, De Marco, \& Barlow 1998, see their Figure 6). For 2MASS J13125770$-$6240599, the relative strengths of these emission lines implies a subtype of WC8. By comparison, the emission lines in the spectrum of 2MASS J18281180$-$1025424 are much weaker relative to the continuum, and there is practically no sign of He or O {\sc v}.  The EW ratio of  C {\sc iv} ($\lambda$5801~\AA) and C {\sc iii} ($\lambda$5696 ~\AA), with a measured value of 0.4, and the lack of O {\sc v} emission, is consistent with WC9 spectral type. 

Although the emission-line EW ratios of 2MASS J18281180$-$1025424 are consistent with WC9 classification, the individual EWs have only $\sim$10\% the strength of what is normally observed for single WC9 stars (e.g., see Conti \& Massey 1989).  Unusually weak emission lines have been observed in the spectra of binary systems containing WCL stars, such as the WC9+B0{\sc I} binary WR 70, whose emission spectrum is diluted by the continuum flux of the bright supergiant companion. In addition to the unusually bright continuum flux in the spectrum of 2MASS J18281180$-$1025424, there is a relatively narrow emission line of H{ \sc i} at $\lambda$6563 \AA, superimposed on the broad He {\sc ii} emission line of the WC9 star, and a weak He {\sc i} absorption line near $\lambda$8680 \AA (Figure 3, \textit{right panel}).  These features are a characteristic of late-O and early-B giants and supergiants. Thus, we interpret the detection of these spectral features, and the continuum-diluted C emission lines, as strong evidence for a luminous OB supergiant companion to the WC9 star in 2MASS J18281180$-$1025424. Moreover, as we will see in \S5.1, 2MASS J18281180$-$1025424 is a hard X-ray source, providing strong evidence for colliding stellar winds.

The luminosity class of the OB companion can be constrained photometrically by quantifying the amount of continuum flux necessary to dilute the emission lines of the WC9 star.  The EW of a line is given by $(f_{\textrm{\scriptsize{line}}}-f_{\textrm{\scriptsize{cont}}})d\lambda/f_{\textrm{\scriptsize{cont}}}$, where $f_{\textrm{\scriptsize{line}}}$ and $f_{\textrm{\scriptsize{cont}}}$ are fluxes in the line and continuum, respectively. For a WC9+OB binary system, the total continuum flux is the sum of the individual continuum fluxes of the two components, $f_{\textrm{\scriptsize{cont}}}=f_{\textrm{\scriptsize{cont}}}(\textrm{WC9})+f_{\textrm{\scriptsize{cont}}}(\textrm{OB})$. The EW expressions for 2MASS J18281180$-$1025424 were combined with those of comparison WC9 stars, assuming a universal $f_{\textrm{\scriptsize{cont}}}(\textrm{WC9})$, and then we solved for $f_{\textrm{\scriptsize{OB}}}/f_{\textrm{\scriptsize{WC9}}}$. We calculated this ratio by combining the C {\sc iv} ($\lambda$5801~\AA), C {\sc iii} ($\lambda$5696 ~\AA) and C {\sc iii} ($\lambda$4650 ~\AA) EW measurements of 2MASS J18281180$-$1025424 with those of WR 81, WR 88, and WR 92, which are putatively single, non-dusty WC9 stars, according to van der Hucht et al. (2001). Averaging the results, we obtain $f_{\textrm{\scriptsize{OB}}}/f_{\textrm{\scriptsize{WC9}}}=4.8\pm1.5$ and a corresponding brightness difference of $\Delta m=1.7\pm0.8$ between the WC9 star and the companion. Assuming $M_V=-4.6\pm0.4$ mag for WC9 stars (van der Hucht 2001), we obtain $M_V=-6.3\pm0.9$ mag for the OB companion, which places it in the luminosity class of supergiants, such as the B0 {\sc I}a stars HD 91949, HD 94909, and HD 122879, which all have $M_V=-6.4$ mag (Crowther 2006a). Thus, based on the available data, we classify 2MASS J18281180$-$1025424 as a WC9+OB{\sc I} binary.

\subsection{Near-Infrared Identifications}
Table 4 lists the EWs of prominent WR emission lines detected in the $K$-band. Uncertainties are $\approx$10\% of the EW values and are the the result of imperfect fitting of the spectral lines and noise. We estimated WN and WC subtypes for these sources based on criteria presented in Figer et al. (1997) and Crowther et al. (2006b).  

Figure 4 displays the $K$-band spectra of 4 stars: 2MASS J12285099$-$6317002, 2MASS J12110256$-$6257476, 2MASS J14212314$-$6018041, and 2MASS J15352652$-$5604123; exhibiting strong emission lines of He and N, characteristic of WN stars. Strong He {\sc ii} emission at $\lambda$2.189 {\micron} relative to Br$\gamma$$+$He {\sc i} at $\lambda$2.165 {\micron} is consistent with subtypes of WN7 and earlier (Figer et al. 1997, see their Figure 1; Crowther et al. 2006b, their Figure 5). For 2MASS J12285099$-$6317002, the ratio of He {\sc ii} ($\lambda$2.189 {\micron}) to Br$\gamma$$+$He {\sc i} ($\lambda$2.165 {\micron}) is high relative to the other stars, and is consistent with subtypes of WN3--6, while the ratio of He {\sc ii} ($\lambda$2.189 {\micron}) to He {\sc i}$+$N {\sc iii} ($\lambda$2.115 {\micron}) is consistent with WN4 subtype. For early WN stars, an additional diagnostic is provided by the ratio of N {\sc v} ($\lambda$2.110 {\micron}) to He {\sc i}$+$N {\sc iii} ($\lambda$2.115 {\micron}) (Crowther et al. 2006b);  N {\sc v} is relatively strong in WN3--4 stars, while He {\sc i}$+$N {\sc iii} is relatively strong in WN5--6 stars. Based upon these diagnostic criteria, we classify 2MASS J12285099$-$6317002, 2MASS J12110256$-$6257476,  2MASS J14212314$-$6018041, and 2MASS J15352652$-$5604123 as WR stars of subtype WN4--5, WN6, WN6, and WN7, respectively. The WN star 2MASS J12121681$-$6246145 is displayed in Figure 5, exhibiting emission lines that are somewhat weak compared with the WN stars in Figure 4. N {\sc v} has comparable strength to N {\sc iii}/He {\sc i}, characteristic of WN4--5 stars, while Br$\gamma$$+$He {\sc i} and He {\sc ii} are comparable in strength, consistent with WN5--6 classification. Therefore, we classify 2MASS J12121681$-$6246145 as having subtype in the range WN4--6.

Figure 6 contains the spectra of 3 WN stars, 2MASS J16441069$-$4524246, 2MASS J16465342$-$4535590 and 2MASS J15595671$-$5159299, which were observed in poor weather conditions. Cloudy conditions interrupted our observations, which prevented the observation of a standard star, so no telluric correction was applied to the spectra in Figure 6. Nonetheless, the major WN diagnostic features were detected, so the WN subtype can still be approximated. For 2MASS J16441069$-$4524246, the ratios of EWs for He {\sc ii} ($\lambda$2.189 {\micron}) to Br$\gamma$$+$He {\sc i} ($\lambda$2.165 {\micron}), and He {\sc ii} ($\lambda$2.189 {\micron}) to He {\sc i}$+$N {\sc iii} ($\lambda$2.115 {\micron}), indicate a subtype in the range WN7--8, which is consistent with our optical classification for this star (see \S3.1). The emission-line EW ratios of 2MASS J16465342$-$4535590 also indicate a WN7--8 subtype. For 2MASS J15595671$-$5159299, the emission lines appear very broad, and He {\sc ii} ($\lambda$2.189 \micron) dominates Br$\gamma$$+$He {\sc i}. The relative line strengths in this star are consistent with the WN5--6 subtype. Furthermore, the FWHM of the He {\sc ii} ($\lambda$2.189 $\micron$) line is 285 ~\AA~for this star; according to Crowther et al. (2006b), stars with FWHM of He {\sc ii} ($\lambda$2.189 \micron) $\ge$ 130 ~\AA are classified as broad-lined WRs. Thus we classify 2MASS J15595671$-$5159299 as a WN5--6b star.

The $K$-band spectrum of 2MASS J12100795$-$6244194 is displayed in Figure 7, and exhibits very strong C {\sc iv} emission at $\lambda$2.076 $\micron$, in addition to C {\sc iii} emission at $\lambda$2.110 ${\micron}$. These are typical features of WC stars.  There is also a weaker signature of He {\sc ii} at $\lambda$2.189 {\micron}, and a relatively narrow and weak line of Br$\gamma$, which may be nebular contamination, as WC stars are completely devoid of H. The ratio of the EW of C {\sc iv} to that of C {\sc iii} can be used as a WC subtype diagnostic (Figer et al. 1997, Crowther et al. 2006b).  The large ratio for 2MASS J12100795$-$6244194 indicates a subtype of WC5--6.

\section{Extinction and Distance}
The photometry of each WR is presented in Table 2. Using our classification of WR subtypes, we can use the colors and absolute magnitudes of known WR stars of the same subtype to compare with our sample. Once an estimate of the extinction along the line-of-sight to these stars is obtained, an approximate distance can be derived. We adopted intrinsic $J-K_s$ and $H-K_s$ colors and $M_{K}$ values by comparison with known WRs from Crowther et al. (2006b, their Table A1), and used the extinction ratios for the $JHK_s$ bands from Indebetouw et al. (2005). We calculated $A_{K_s}$ using the $J-K_s$ and $H-K_s$ colors separately and averaged the two results, for a final $A_{K_s}$ estimate. Assuming $M_{K}$ based upon WR subtype, we calculated the heliocentric distance to each WR star, $R$, in kpc. Assuming a distance of 8 kpc to the Galactic center (Reid 1993), we also calculated the galactocentric distance of each star, $R_{G}$, in kpc. The results are listed in Table 5. The $A_{K_s}$ values of these stars lie between $\approx0.8$--1.6 mag. 

For the WC9+OB{\sc I} binary 2MASS 18281180$-$1025424, we compute a total system brightness of $M_V=-6.5\pm0.9$ mag, using the individual magnitudes of the components (\S3.1). According to the Naval Observatory Merged Astronomical Dataset (NOMAD\footnote{http://www.navy.mil/nomad.html}), 2MASS J18281180$-$1025424 has $V\simeq14.83$ mag and $B-V\simeq2.30$ mag. According to van der Hucht (2001), WC9+OB {\sc I} binaries are assigned $(B-V)_0= -0.45$ mag, which implies $E(B-V)\simeq2.75$ mag for 2MASS J18281180$-$1025424. Using the relation $A_V/R=E(B-V)$, where $R=3.1\pm0.1$ (Cardelli et al. 1989), we compute an extinction of $A_V=8.5\pm0.1$ mag. Thus, we calculate a distance modulus to the 2MASS J18281180$-$1025424 system of $12.8\pm0.9$ mag, which corresponds to a distance of $3.7\pm1.5$ kpc. 

Figure 8 illustrates the galactocentric distribution of the WR stars identified in this work, as well as those from Hadfield et al. (2007). Our observations mainly targeted sources in the southern sky, near Galactic longitudes of 300$^{\circ}$ and 330$^{\circ}$. Most of the WR stars confirmed here lie at distances of several kpc or less, generally consistent with the distribution of WRs in van der Hucht (2001, 2006). Exceptions are 2MASS J12285099$-$6317002 and 2MASS J15352652$-$5604123, which appear to lie at heliocentric distances of 14.1 and 9.8 kpc, respectively. 

\section{Discussion}
\subsection{X-ray Emission from 2MASS J18281180$-$1025424}
The position of the WC9+OB{\sc I} binary 2MASS J18281180$-$1025424 is coincident with the X-ray source XGPS-I J182811$-$102540 (Hands et al. 2004), which was observed on 2001, March 26 during a survey of the Galactic plane by \textit{XMM-Newton} (\textit{XMM}) with the EPIC instrument (Bignami et al. 1990). Figure 9 compares the 2MASS and \textit{XMM} fields containing this star. The position of  XGPS-I J182811$-$102540 (R.A.=18h28m11.92s, Dec=$-$10d25m40.9s, J2000) is separated from the infrared source by $2\farcs29$. The X-ray source has a flux of $(3.5\pm1.6)\times10^{-3}$ counts s$^{-1}$ within an energy range of 0.4--6.0 keV, as measured by the pn CCD of the EPIC instrument. The flux detected within the 2.0--6.0 keV  hard band is a factor of 2 higher than the flux detected within the 0.4--2.0 keV soft band. This information may be used to constrain the spectral properties of the X-ray source by calculation of the hardness ratio, which is defined as $HR=(H-S)/(H+S)$, where $H$ and $S$ are the fluxes in the hard and soft energy bands, respectively.  Sources were classified in Hands et al. (2004) as soft, medium, or hard, if they had hardness ratios of $HR<-0.5$, $-0.5<HR<0.5$, or $HR>0.5$, respectively. The source XGPS-I J182811$-$102540 was reported having $HR=0.65$ and, thus, is classified as a hard X-ray source. The relatively high $HR$ value may be due to several causes. Soft X-rays suffer significant absorption by interstellar gas and dust, so the intrinsic X-ray spectrum may in fact be softer than the hardness ratio implies. Nonetheless, the detection of hard X-rays in the range 2--6 keV indicates the presence of a hot component ($T> {\rm few}\times10^7$ K). Such high plasma temperatures are not a ubiquitous characteristic of single WR stars of any subtype, and the detection of X-rays from WC9 single stars is uncommon (Skinner et al. 2006).  However, hard X-ray emission is common if WRs are members of binaries, such as $\gamma^2$ Vel and WR 65 (Henley et al. 2005; Oskinova \& Hamann 2008), which are systems in which the supersonic wind of the WR collides with the opposing wind of a companion. The optical spectrum of 2MASS J18281180$-$1025424, presented in \S3.1, indicates the presence of a OB{\sc I} companion to the WC9 star, so the emission of hard X-rays from colliding winds would not be surprising. 

To calculate the X-ray luminosity of 2MASS J18281180$-$1025424 we converted the value of $A_V=8.5$ mag (see \S4) to an equivalent hydrogen column density, using the relation of Predehl \& Schmitt (1995), and obtained $N_{\textrm{\tiny{H}}}=1.5\times10^{22}$ cm$^{-2}$.  We then used \program{WebPIMMS}\footnote{http://heasarc.nasa.gov/Tools/w3pimms.html} to obtain the unabsorbed X-ray flux from the source by entering the observed number of counts in the 0.4--6 keV band and applying a Raymond-Smith thermal plasma spectrum with solar metallicity (Raymond \& Smith 1977). We performed several trials using plasma temperatures in the range $kT=1.5$--3.0 keV, which is a reasonable range of temperatures one could assume for the hot component of colliding-wind binary. The calculations yielded unabsorbed fluxes between ($7.9$--$7.1)\times10^{-14}$ erg cm$^{-2}$ s$^{-1}$, respectively. Thus, we used an unabsorbed flux value of $(7.5\pm0.4)\times10^{-14}$ erg s$^{-1}$ and applied our derived distance of $3.7\pm1.5$ kpc to obtain an X-ray luminosity of  $L_{\textrm{\tiny{X}}}=(1.2\pm1.0)\times10^{32}$ erg s$^{-1}$. This is similar to the X-ray luminosities of known WC9+OB binaries, such as WR 65 (Oskinova \& Hamann 2008). 

\subsection{The Spectral Energy Distribution of 2MASS J18281180$-$1025424}
Most known WC9+OB binaries are associated with excess thermal emission from hot dust, which supplements the free-free excess from the stellar wind (Cohen et al. 1975; WIlliams et al. 1987; van der Hucht et al. 1996; Williams et al. 1998). It is thought that the dust is produced via the collision of a carbon-rich WC wind with a hydrogen-rich OB wind (Le Teuff 2002), which mix and form dust within regions of the collision wake that are shielded from destructive UV photons (e.~g., see Cherchneff et al. 2000; Crowther et al. 2003, and references therein). To investigate the possibility of thermal dust emission from the new WC9$+$OB{\sc I} binary star 2MASS J18281180$-$1025424, we compared the photometry of this source with that of several known WC9 stars from literature, including the dusty WC9d star WR 59 (van der Hucht 2001) and the non-dusty WC9 stars HDM6 and HDM13 (Hadfield et al. 2007).  Figure 10 displays the infrared spectral energy distributions (SED) of these stars, which were de-reddened using the extinction relations of Cardelli et al. (1989) and Indebetouw et al. (2006).  For comparison, Rayleigh-Jeans (RJ) curves of sources with $T_{eff}=40$ kK are also plotted in Figure 10 and scaled to the $J$-band data point of each source, although using any of the full range of known WR/OB temperatures would result in a negligible difference, owing to the temperature degeneracy of the RJ slope. The SEDs of 2MASS J18281180$-$1025424 and the WC9d star WR 59 both rise between the $H$ and $K_s$ bands and exhibit relatively high excess in the mid-infrared (saturated at $\lambda$4.5 {\micron}).  These features are typical of thermal emission from hot dust. By comparison, the smaller excess in the SEDs of HDM6 and HDM13, which maintain a decreasing slope with wavelength, are typical of free-free emission from non-dusty WC9 stars (e.~g., see Cohen et al. 1975). This indicates that the WC9$+$OB{\sc I} binary 2MASS J18281180$-$1025424 is very likely to be a source of thermal emission from hot dust.

In general, the SEDs of dust-producing WC9+OB binaries are modeled with multiple hot dust shells that have a gradient in temperature (Williams, van der Hucht \& Th\'{e} 1987). Detailed SED modeling is beyond the scope of this paper. However, it is reasonable to assume that the large excess in the $K$ band and the overall behavior of mid-infrared SED suggest a dust temperature on the order of $\sim10^3$ K, which is typical of dust-producing WC9+OB binaries. In conclusion, we classify 2MASS J18281180$-$1025424 as a dust-producing WC9d$+$OB{\sc I} binary.

\subsection{Origin of New WR Stars}
Figures 9 and 11 contain 2MASS $K_s$-band field images of the new WRs. In most cases, they appear relatively isolated, with no obvious stellar density enhancement in their vicinity. An exception is the WC8 star 2MASS J13125770$-$6240599. This star is appears associated with the H {\sc ii} region G305 and the star cluster Danks 1 (Georgelin et al. 1988; Danks et al. 1984). This cluster is one of a pair of dense stellar clusters, Danks 1 and 2, that lie near the center of a cavity that is defined by a three-lobed structure of mid-infrared emission. One WR has already been associated with this region, the WC star WR 48a (see Clark \& Porter 2004, and references therein). The system lies at a distance of 3.5--4 kpc from the Sun, which is consistent with our derived photometric distance of 3.6 kpc to 2MASS J13125770$-$6240599. The presence of 2 WC stars within this system suggests that there may be additional WRs in this environment at earlier evolutionary stages, such as WN stars, that have yet to be identified. Within a angular radius of $\approx5{\arcmin}$ surrounding the Danks 1 and 2 clusters, there are 6 additional WR candidates that were selected by our criteria, although they are fainter than the confirmed WC stars in this region by $\approx2$--3 mag in the $K_s$ band and lie in the outskirts of the clusters.  As in Hadfield et al. (2007), our selection criteria are restricted to sources that have only the highest photometric quality in the 2MASS catalog.  As a result, we are biased against the detection of WRs in regions of significant stellar confusion, such as the interior regions of the clusters Danks 1 and 2. Thus, it would not be surprising if, in the future, more WRs are confirmed within Danks 1 and 2 that were not selected by our method. Indeed, if we relax the near-infrared photometric-quality restriction we obtain a total of 17 unidentified WR candidates that appear spatially concentrated in the G305 region, within $\approx5{\arcmin}$ of the clusters. Alternatively, the 2 confirmed WC stars in this region may represent the few remaining WRs from Danks 1, which would be consistent with the upper limit on the age of Danks 1 (5 Myr; Clark \& Porter 2004). If this is the case, the remaining WR candidates in this region will probably be non-WR emission-line stars, similar to those that dominate our sample (e.~g., see Hadfield et al. 2007, their Figure 7). In the future we plan to obtain spectra of all WR candidates in the G305 region, including those having relatively low photometric quality. This will not only elucidate the nature of the Danks 1 and 2 clusters, but will also be useful in quantifying the limitations of our WR selection criteria in regions of high stellar confusion. 

In comparison with 2MASS J13125770$-$6240599, the rest of the WRs appear relatively isolated, as is also the case for most of the 15 WRs reported in Hadfield et al. (2007). This motivates the question as to whether WRs form exclusively in clusters and associations, or may also form via an alternate mode of isolated massive-star formation. It is also possible that WRs form in clusters and associations, but are sometimes dynamically ejected, either by inter-cluster dynamics or the supernova explosion of an erstwhile companion. Indeed, it has been noted that half of all O stars are runaways (de Wit et al. 2004, 2005, and references therein), and the same might be true for WRs, since they are descended from O stars (Dray et al. 2005). Any of these possibilities could be an explanation for the spatial distribution of our sample. Proper motion measurements would be required to constrain these origin scenarios, in addition to a more complete spectroscopic survey of neighboring stars. We save a full investigation of the environments of these stars for a future paper.

\section{Conclusions}
The selection of candidate WRs based upon their near-infrared and mid-infrared colors has proven successful. The method mainly benefits from the mid-infrared excess emission, generated via free-free reactions in the dense, ionized winds of WRs. We identified 12 new WRs, including 9 WN and 3 WC stars. These discoveries add to the earlier results of Hadfield et al. (2007), and bring the total number of WRs identified in this way to 27. However, the modest detection rate of $\approx$10\% for WRs is accompanied by a non-WR, emission-line star detection rate of $\approx65\%$, so the latter are our main contaminant. In the interest of increasing the efficiency of WR detection, we must consider ways in which our selection criteria may be improved. In principle, we may exploit the fact that the free-free emission spectra of WRs tends to become optically thick at wavelengths longer than $\approx$5 {\micron} (Wright \& Barlow 1975; Cohen et al. 1975), with a predictable spectral index. This information may be helpful in distinguishing true WRs in our sample from other objects that exhibit similar color excesses in the 2MASS and \textit{Spitzer}/IRAC bands. We are also in the process of characterizing the spectral energy distributions of known WRs out to 24 {\micron}, with the inclusion of data from the \textit{Spitzer}/MIPSGAL survey of the inner Galactic plane (Carey et al. 2006). Although a more thorough analysis of the 24 {\micron} photometry of WRs is forthcoming in an infrared phenomenological study of WRs, our preliminary results are encouraging:  we have determined that confirmed, non-dusty WRs occupy a narrow subset of the 8--24 {\micron} color space spanned by our current sample of WR candidates. Although a refinement of our selection criteria using 8--24 {\micron} colors may be biased toward WNs, we speculate that WC9 stars with hot dust may also be distinguishable from other, cooler dusty sources whose excesses continue to rise with increasing wavelength beyond  $\lambda20$ {\micron}. In any case, assuming that our evolving WR selection criteria maintains a static success rate of 10\%, we may expect to increase the known WR population in the Galaxy from $\sim$300 to at least $\sim$800, and perhaps over 1000 once we pursue candidate WRs within the GLIMPSE-II survey of the inner Galaxy. This number is more-or-less consistent with the presumed number of Galactic WRs that are expected to have $K\lesssim14$ mag (Shara et al. 1999), within the 2MASS sensitivity limit.  Thus, our WR survey using 2MASS$+$\textit{Spitzer}/GLIMPSE has the potential to complete the WR sample for the near-half of the Galaxy within the \textit{Spitzer}/GLIMPSE survey area. A sample increase of this significance will be invaluable in our effort to understand the evolution of WRs and their Galactic demography, the evolution of massive stars in general, and the recent massive star formation history of the Galaxy. 

\begin{acknowledgments}
This publication makes use of data products from the Two-Micron All-Sky Survey (2MASS), which is a joint project of the University of Massachusetts and IPAC/Caltech, funded by the NASA and the NSF. The research was based on observations made with the \textit{Spitzer Space Observatory}, Palomar Observatory, Cerro Tololo International Observatory, and the Anglo Australian Observatory. We are thankful to the referee for providing insightful comments and valuable suggestions for improvement. We thank L.~J. Hadfield, J.~D. Smith, and A.~P. Marston for their contributions to our search for Galactic WRs. We also thank Jean Mueller for her assistance with the Palomar observations.

\end{acknowledgments}

\begin{deluxetable}{lccccc}
\tablecolumns{4}
\tablewidth{0pc}
\tabletypesize {\scriptsize}
\tablecaption{Spectroscopic Observations of WRs}
\tablehead{
\colhead{Star} & \colhead{Observation Date} & \colhead{Telescope/Instrument} & \colhead{$R$}  \\
\colhead{2MASS ID} & \colhead{(UT)} & \colhead{} & \colhead{} 
}
\startdata
J11555211$-$6245022   & 2008-07-25    & CTIO 4 m/RCSpec   & 1500 \\ 
J12100795$-$6244194   & 2008-05-17    & AAT/IRIS2        & 2400 \\ 
J12110256$-$6257476   & 2008-05-17    & AAT/IRIS2    & 2400 \\ 
J12121681$-$6246145 & 2006-07-20    & SOAR/OSIRIS    & 1200 \\ 
J12285099$-$6317002   & 2008-05-14    & AAT/IRIS2      & 2400 \\ 
J13125770$-$6240599   & 2006-03-20    & CTIO 4 m/RCSpec      & 1500   \\ 
J14212314$-$6018041   & 2008-05-16    & AAT/IRIS2    & 2400 \\ 
J15352652$-$5604123   & 2008-05-17    & AAT/IRIS2    & 2400 \\ 
J15595671$-$5159299   &    2008-07-28  & SOAR/OSIRIS & 1200 \\ 
J16441069$-$4524246   &   2008-07-25   & CTIO 4 m/RCSpec    & 1500 \\ 
J16441069$-$4524246   &   2008-07-28   & SOAR/OSIRIS    & 1200 \\ 
J16465342$-$4535590   & 2008-07-28    & SOAR/OSIRIS   & 1200 \\ 
J18281180$-$1025424   & 2003-03-06    & Hale 5 m/DBSpec   & 2000 
\enddata
\end{deluxetable}

\setlength{\tabcolsep}{0.06in}
\renewcommand{\arraystretch}{1.05}
\begin{landscape}
\begin{deluxetable}{lrrcrrrrrrr}
\tablecolumns{11}
\tablewidth{0pc}
\tabletypesize {\scriptsize}
\tablecaption{Coordinates and Photometry for New WRs}
\tablehead{
\colhead{2MASS}& \colhead{$l$} & \colhead{$b$} &\colhead{$R$} &  \colhead{$J$} & \colhead{$H$} & \colhead{$K_s$} &  \colhead{$M_{3.6}$}  & \colhead{$M_{4.5}$}  & \colhead{$M_{5.8}$} &\colhead{$M_{8.0}$} \\ [2pt]
\colhead{Designation} & \multicolumn{2}{c}{(deg, J2000)}& \colhead{(mag)} & \colhead{(mag)} & \colhead{(mag)} & \colhead{(mag)} & \colhead{(mag)} & \colhead{(mag)} & \colhead{(mag)} & \colhead{(mag)}}
\startdata
J11555211$-$6245022 &  296.620230    &   $-$0.561908    & 14.03       & $11.65\pm0.03$ &$10.76\pm0.03$&$10.14\pm0.03$ & $9.33\pm0.04$ &$8.94\pm0.05$&$8.70\pm0.04$& $8.33\pm0.03$  \\                                

J12100795$-$6244194 &  298.222207    &   $-$0.245330     & \nodata    & $12.75\pm0.02$ & $11.08\pm0.02$ & $9.74\pm0.02$ & $8.29\pm0.05$ & $7.85\pm0.04$ & $7.52\pm0.04$ &  $7.36\pm0.04$ \\               

J12110256$-$6257476 &  298.360150    &   $-$0.450644   &  \nodata   & $12.02\pm0.03$ & $10.70\pm0.03$ & $9.85\pm0.03$   &$8.98\pm0.04 $ & $8.62\pm0.05$ & $8.37\pm0.04 $ & $8.07\pm0.03 $\\                 

J12121681$-$6246145 &  298.469979    &   $-$0.238728   &  \nodata   & $12.07\pm0.04$ & $10.72\pm0.03$ & $9.88\pm0.03$ & $9.08\pm0.03$ & $8.68\pm0.05$ & $8.39\pm0.04 $ & $8.10\pm0.02 $ \\           
 
J12285099$-$6317002 &  300.396357   &    $-$0.522936   &  \nodata   & $13.56\pm0.03$ & $12.55\pm0.03$  & $11.84\pm0.03$ & $11.00\pm0.04$ & $10.59\pm0.06$ & $10.32\pm0.04$ & $10.05\pm 0.04$ \\      

J13125770$-$6240599 &  305.398457    &       0.085275       &  15.12      & $10.83\pm0.03$ &  $9.83\pm0.03$ & $8.98\pm0.03$ & $8.37\pm0.04 $ & $7.94\pm0.05 $ & $7.77\pm0.04$ & $7.46\pm0.03$\\                                   

J14212314$-$6018041 &    313.855830    &        0.648914     &  \nodata   & $11.68\pm0.02$ & $10.43\pm0.02$ & $9.62\pm0.02$ & $8.96\pm0.03$ & $8.53\pm0.06$ & $8.23\pm0.04$ & $7.94\pm0.03 $\\ 

J15352652$-$5604123 &    324.413864  &   $-$0.196269     &  \nodata   & $13.84\pm0.04$ & $12.39\pm0.04$ & $11.46\pm0.03$ & $10.58\pm0.05 $ & $10.12\pm0.06 $ & $9.84\pm0.05 $ & $9.49\pm0.03$ \\ 

J15595671$-$5159299 &   329.760410   &      0.778750       &   \nodata  & $10.88\pm0.03$& $9.52\pm0.03  $ &$8.66\pm0.03$ & $7.83\pm0.06 $ & $7.43\pm0.06$ & $7.17\pm0.04$ & $6.78\pm0.03$ \\ 
 
J16441069$-$4524246 &    339.558159  &         0.268472   &  15.26        & $10.06\pm0.02$ & $9.00\pm0.03 $ & $8.29\pm0.02$ & $7.44\pm0.06$& $6.99\pm0.07$& $6.84\pm0.03$ & $6.45\pm0.03$  \\

J16465342$-$4535590 &    339.721124  & $-$0.217803      &   \nodata  & $10.75\pm0.02$ & $9.53\pm0.03$ & $8.73\pm0.02 $ & $7.99\pm0.04$ & $7.52\pm0.05$ & $7.28\pm0.03$ & $6.90\pm0.03$ \\

J18281180$-$1025424 &    21.014727     & 0.347933    & 13.4         &  $8.36\pm0.03$  & $6.93\pm0.04$ &  $5.63\pm0.03$ & $4.50\pm0.09$  &       $<$6.5           &   $3.54\pm0.06$  & $2.66\pm$0.04\tablenotemark{a}  

\enddata
\tablenotetext{a}{8 {\micron} flux from the MSX Point Source Catalog, assuming the zeropoint flux of 58.4 Jy from Cohen, Hammersley, \& Egan (2000).}
\end{deluxetable}
\end{landscape}

\begin{landscape}
\begin{table*}
\small
\caption{Equivalent widths (W$_{\lambda}$) and FWHM
  (both in ~\AA) of prominent, optical emission lines.  WR
  subtypes have been assigned using the classification schemes of
  Smith et al. (1996) for WN stars, and Crowther, De Marco \& Barlow (1998) for WC stars. The measurements are accurate to within 10\% of their value.}
\begin{tabular}{cc@{\hspace{1mm}}cc@{\hspace{1mm}}cc@{\hspace{1mm}}cc@{\hspace{1mm}}cc@{\hspace{1mm}}cc@{\hspace{1mm}}cc@{\hspace{1mm}}cl}
\hline Star & \multicolumn{2}{c}{N\, {\sc iii}/C {\sc iii}} & \multicolumn{2}{c}{He\, {\sc ii}}&
\multicolumn{2}{c}{He\, {\sc ii}}
&\multicolumn{2}{c}{C {\sc iii}}
&\multicolumn{2}{c}{C {\sc iv}}
&\multicolumn{2}{c}{He\, {\sc i }}
&\multicolumn{2}{c}{H\,$\alpha$}&Spectral\\
2MASS&\multicolumn{2}{c}{(4640/50~\AA)}&\multicolumn{2}{c}{(4686~\AA)}&\multicolumn{2}{c}{(5411~\AA)}&\multicolumn{2}{c}{(5696~\AA)}&\multicolumn{2}{c}{(5801~\AA)}&\multicolumn{2}{c}{(5876~\AA)}&\multicolumn{2}{c}{(6560~\AA)}&Type\\
Designation&FWHM&W$_{\lambda}$&FWHM&W$_{\lambda}$&FWHM&W$_{\lambda}$&FWHM&W$_{\lambda}$&FWHM&W$_{\lambda}$&FWHM&W$_{\lambda}$&FWHM&W$_{\lambda}$&\\
\hline 
J11555211$-$6245022 & 21 & 39 & 19 & 153 & 20 & 27 & \nodata & \nodata & 34 & 12 & 21 & 10 & 29 & 108 & WN6\\
J16441069$-$4524246 & 15 & 46  &  \nodata & \nodata  & 24 &  25  &  58 & 28 &  23 & 4 &  32 & 67 & 38 & 75 & WN8  \\ 
J13125770$-$6240599 & 35 & 396 & 50 & 240 & 40 & 27 &                     67 & 497 & 41 & 435 & 48 & 127 & 71 & 154 & WC8 \\
J18281180$-$1025424{\tablenotemark{a}} & 31 &16   & 19  & 6      &  \nodata & \nodata & 43 &  30 & 36 & 12 & weak & weak & 48 & 24 & WC9d+OB{\sc I}\tablenotemark{b} \\
\hline
\tablenotetext{a}{For 2MASS 13125770-6240599, we also detect O {\sc v} emission at 5590~\AA, with a FWHM and EW  of 47 and 25~\AA, respectively.}
\tablenotetext{b}{See \S5.2 for a justification of the WC9d classification.}
\end{tabular}
\label{tab:prop}
\end{table*}
\end{landscape}

\begin{landscape}
\begin{table*}
\small
\caption{Equivalent widths (W$_{\lambda}$, in ~\AA) of
  near-IR emission-lines used to classify WRs.  WR subtypes have been assigned using the classification
  scheme of Crowther et al. (2006b) and Figer et al. (1997). The measurements are accurate to within 10\% of their value.}
\begin{center}
\begin{tabular}{cccccccl}
\hline
Star&C\, {\sc iv}&C\, {\sc iii}&N\, {\sc v}&N\, {\sc iii}/He\, {\sc ii}&He\, {\sc ii}+Br$\gamma$&He {\sc ii}&Spectral\\
2MASS ID&(2.076~$\mu$m)&(2.110~$\mu$m)&(2.110~$\mu$m)&(2.115~$\mu$m)&(2.165~$\mu$m)&(2.189~$\mu$m)&Type\\
\hline
J12110256$-$6257476&  \nodata & \nodata  & 3 & 10  & 30 & 30  & WN6  \\
J12121681$-$6246145&  \nodata & \nodata & 3 & 9 & 30  & 30  & WN4--6  \\
J12285099$-$6317002&  \nodata&\nodata   & 9 & 20   & 30  & 123  & WN4--5  \\
J14212314$-$6018041&  \nodata & \nodata &  6 & 40 & 40 & 30 & WN6  \\
J15352652$-$5604123&  \nodata & \nodata  & 6 & 20  & 44 & 39   & WN7  \\
J15595671$-$5159299&  \nodata & \nodata  & \nodata & 100   & 40  & 140   & WN5--6b  \\ 
J16441069$-$4524246&  \nodata & \nodata  &  \nodata & 70  & 100  & 40  & WN7  \\ 
J16465342$-$4535590&  \nodata & \nodata  & \nodata & 60  & 80  & 40  & WN7 \\ 
J12100795$-$6244194 &   260 & 40  & \nodata & \nodata & 3  & 10   & WC5--7 \\
\hline
\end{tabular}
\end{center}
\label{tab:propir}
\end{table*}
\end{landscape}

\begin{deluxetable}{ccccccccc}
\tablecolumns{9}
\tablewidth{0pc}
\tablecaption{Extinction and Approximate Distances to New WRs}
\tablehead{
\colhead{2MASS}& \colhead{Spectral} & \colhead{$K_s$} &\colhead{$A^{J-K_{s}}_{K_{s}}$} &  \colhead{$A^{H-K_{s}}_{K_{s}}$} & \colhead{$\overline{A_{K_{s}}}$} & \colhead{$M_{K_s}$} &  \colhead{$R$}  & \colhead{$R_G$}  \\ [2pt]
\colhead{Designation} & \colhead{Type} & \colhead{(mag)} & \colhead{(mag)} & \colhead{(mag)} & \colhead{(mag)} & \colhead{(mag)} & \colhead{(kpc)} & \colhead{(kpc)}}
\startdata
J11555211$-$6245022  &	WN6          & 10.14  & 0.89 & 0.85 & 0.87 &-4.41 & 5.4 & 7.8  \\
J12100795$-$6244194  &	WC5--7	  &  9.74   & 1.60 & 1.39 & 1.49 &-4.59 & 3.7 & 7.5 \\
J12110256$-$6257476  &	WN6	  &  9.85   &  1.21 & 1.07 & 1.14 &-4.77 & 5.0 & 7.5 \\
J12121681$-$6246145  &	WN4--6	   &  9.88   & 1.22 & 1.04 & 1.13 &-4.77 & 5.1 & 7.5 \\
J12285099$-$6317002  &	WN4--5  & 11.84  &  0.91 & 0.81 & 0.86 &-4.77 &14.1& 12.3\\
J13125770$-$6240599  &	WC8	  & 8.98    & 0.98 & 0.76 & 0.87 &-4.65 & 3.6 & 7.1\\
J14212314$-$6018041  &	WN6	  & 9.62   &  1.13 & 0.98 & 1.05 &-4.77 & 4.6 & 6.3\\
J15352652$-$5604123  &	WN7	 & 11.46  &  1.35 & 1.20 & 1.27 &-4.77 & 9.8 & 5.7  \\
J15595671$-$5159299  &	WN5--6b	 &  8.66   & 1.24 & 1.08 & 1.16 &-4.77 & 2.8 & 6.2\\
J16441069$-$4524246  &	WN7   	 &  8.29  &  0.94 & 0.80 & 0.87 &-4.77 & 2.7 & 6.0 \\
J16465342$-$4535590  &	WN7 	 & 8.73 & 1.11 & 0.96 & 1.03 &-4.77 & 3.1 & 5.7   \\
J18281180$-$1025424  &	WC9+OB{\sc I}\tablenotemark{a}        & \nodata & \nodata & \nodata & \nodata & \nodata & 3.7\tablenotemark{b} & 4.8
\enddata

\tablenotetext{a}{$M_V=-6.5\pm0.9$ calculated in \S3.1.}
\tablenotetext{b}{Distance calculation for this source is based on optical data and is described in \S4..}
\end{deluxetable}
\clearpage
\newpage

\begin{landscape}
\begin{figure*}[t]
\centering
\epsscale{1}
\plottwo{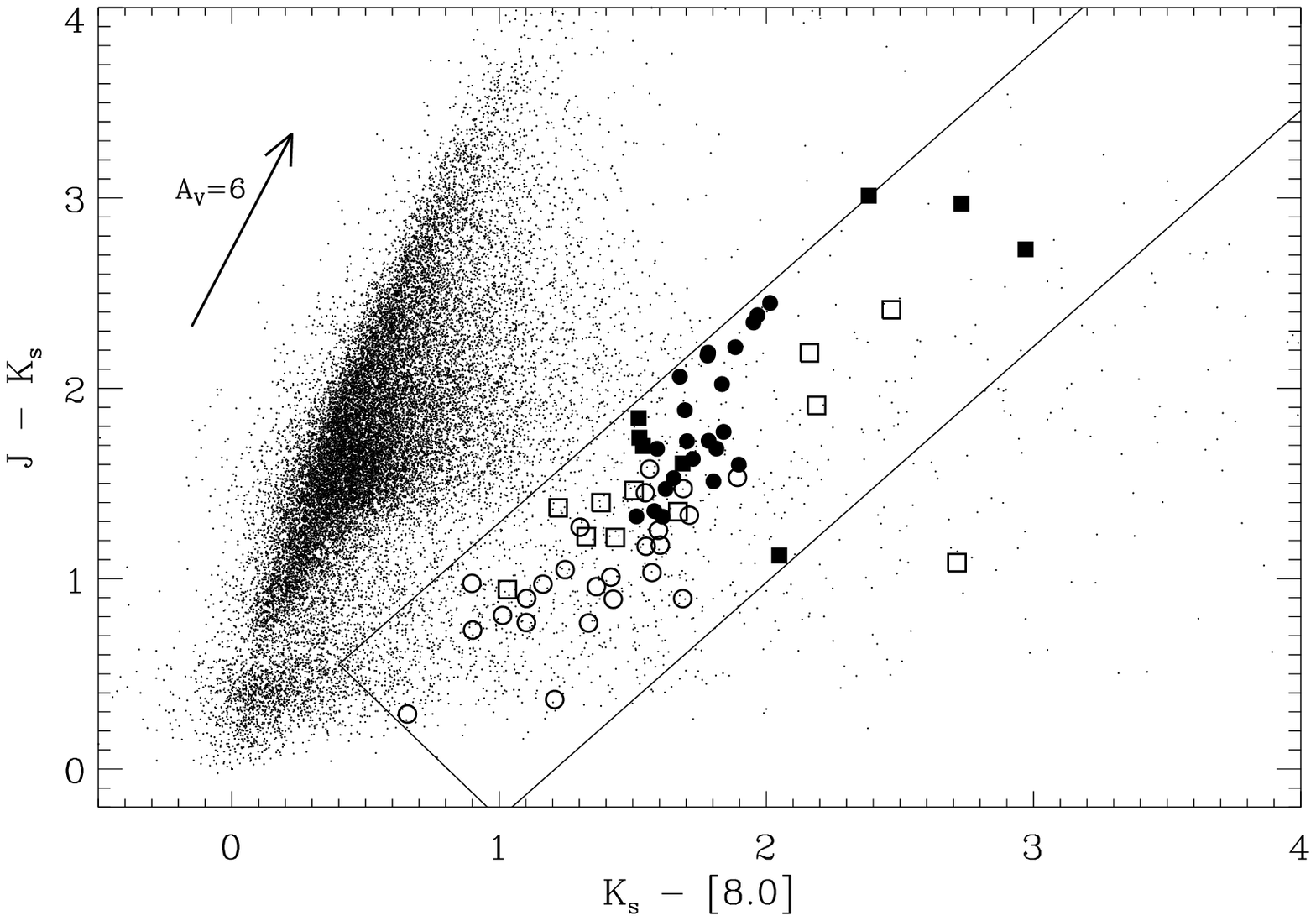}{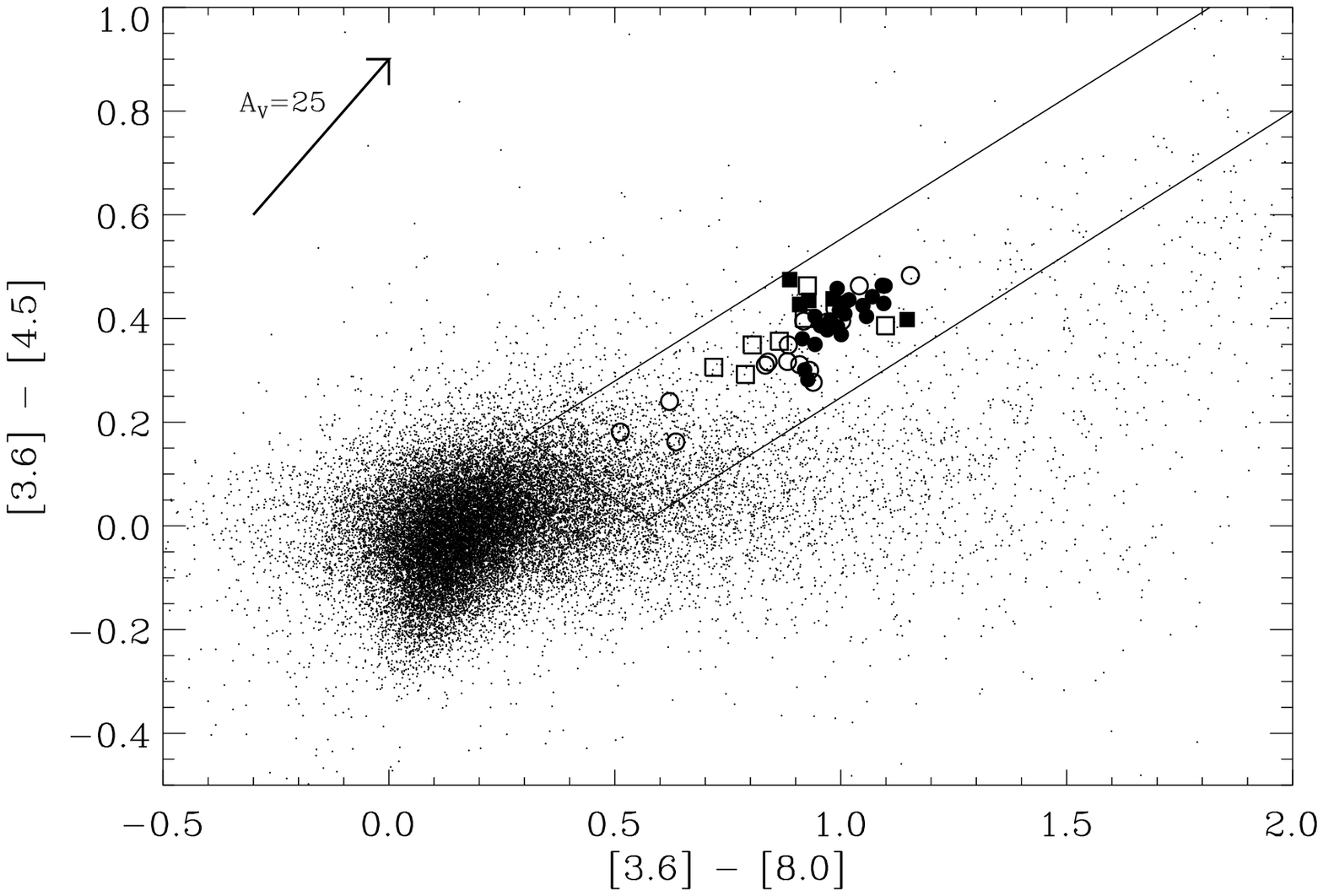}
\caption[]{\linespread{1}\normalsize{Color-color diagrams (magnitude scale) comparing the near-infrared and mid-infrared photometry of WRs to that of $\approx$89000 field stars (\textit{small points}), randomly selected from the \textit{Spitzer}/GLIMPSE-I survey area. WRs are clearly separable from the main locus of field stars in both $K_{s} - [8.0~\micron]$ versus  $J-K_{s}$, and [3.6]-[8.0] versus [3.6]-[4.5]~\micron~color space (left and right panels, respectively). The \textit{solid lines} enclose the WR color spaces used for this work, which were defined using known WRs from van der Hucht (2001; \textit{open symbols}) in the GLIMPSE-I survey area. WC and WN stars are represented by \textit{squares} and \textit{circles}, respectively.  New WRs reported in this work, as well as those discovered in Hadfield et al. (2007) are also plotted (\textit{filled symbols}). The reddening vector (Indebetouw et al. 2005) is represented by the arrow near the top left of each panel.}}
\end{figure*}
\end{landscape}

\begin{figure*}[t]
\centering
\epsscale{1}
\plotone{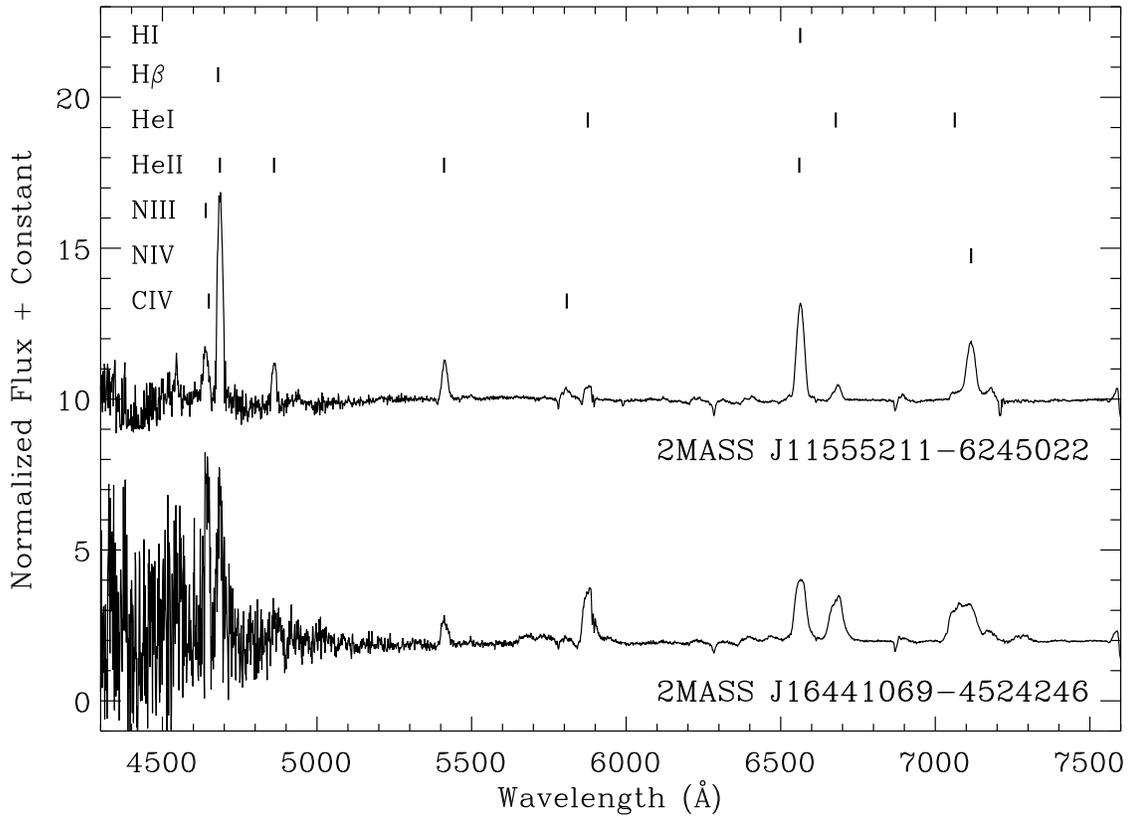}
\caption[]{\linespread{1}\normalsize{Optical spectra of new WRs 2MASS J11555211$-$6245022 (upper spectrum) and 2MASS J16441069$-$4524246 (lower spectrum), obtained with the CTIO 4 m telescope and RCSpec instrument. The emission spectra are dominated by lines of helium and nitrogen, the relative strengths of which are characteristic of nitrogen subtypes WN6 and WN8 for 2MASS J11555211$-$6245022 and 2MASS J16441069$-$4524246, respectively.}}
\end{figure*}

\begin{figure*}[t]
\centering
\epsscale{1}
\plottwo{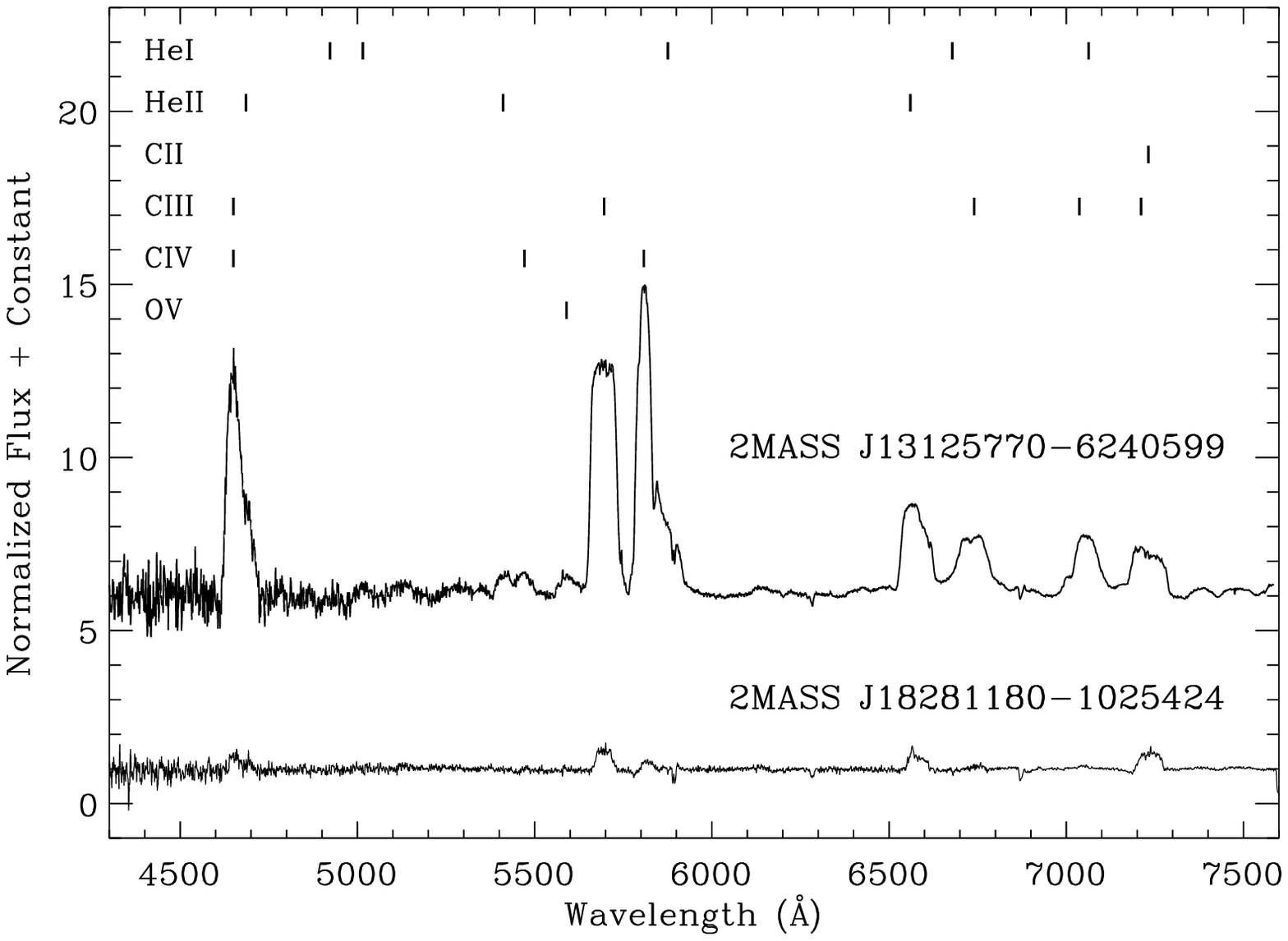}{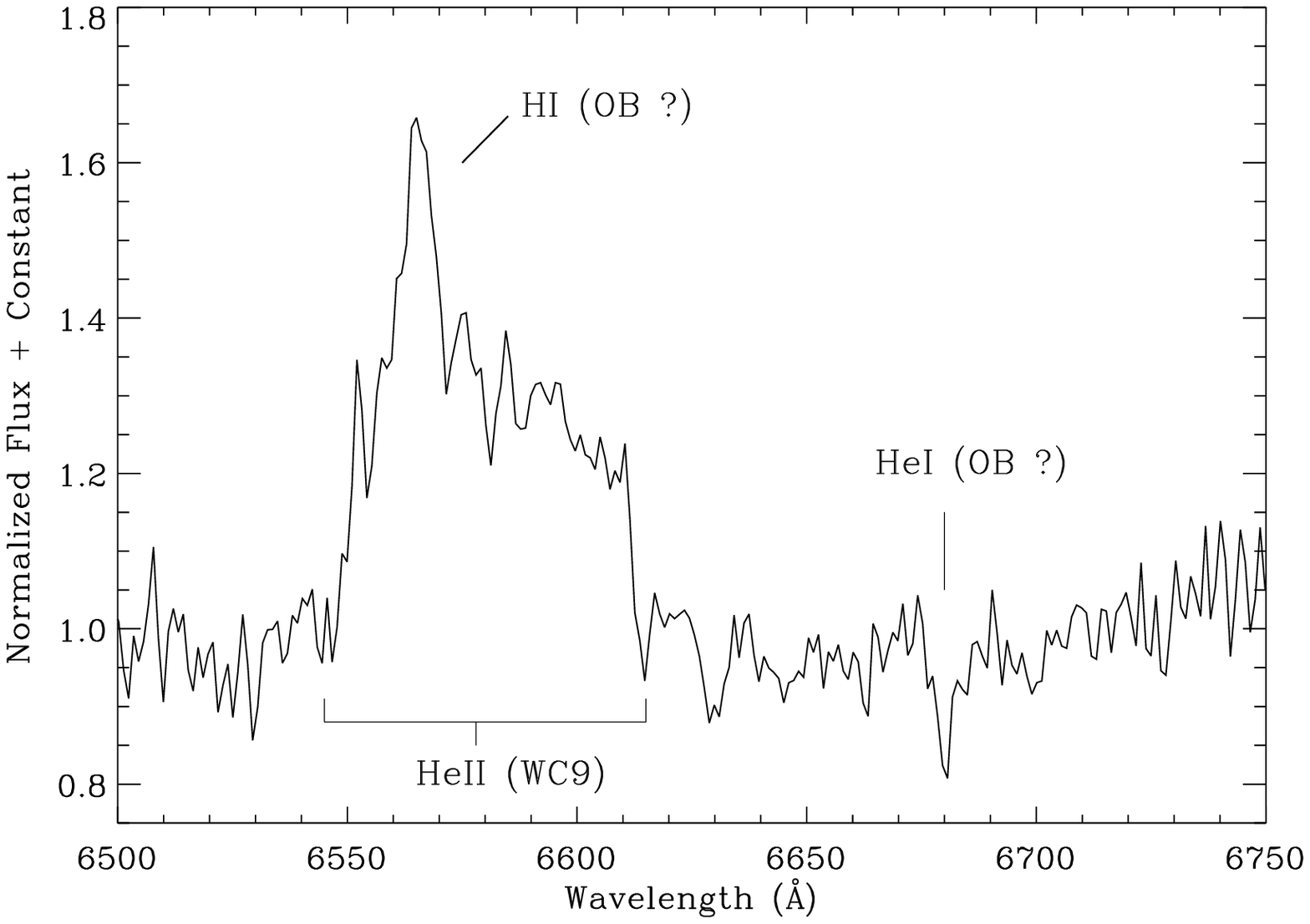}
\caption[]{\linespread{1}\normalsize{\textit{Left panel: }Optical spectra of the new WRs, 2MASS J13125770$-$6240599 and  2MASS J18281180$-$1025424, obtained with the CTIO 4 m/RCSpec and Hale 5 m/DBSpec, respectively. The spectra exhibit emission lines of C {\sc iii} and C {\sc iv}, and weaker lines of He {\sc ii} and O {\sc v}, the relative strengths of which are characteristic of WC8 and WC9 subtypes for 2MASS J13125770$-$6240599 and  2MASS J18281180$-$1025424, respectively.  The weakness of the emission lines in the spectrum of 2MASS J18281180$-$1025424 are likely the result of dilution from the bright continuum of an OB{\sc I} companion. \textit{Right panel :} Expanded view of the spectrum of the weak-lined WC9 star 2MASS J18281180$-$1025424. Relatively narrow H {\sc i} emission at $\lambda$6563 \AA}, superimposed on the broad He {\sc ii} line of the WC9 star, may be the feature of an OB supergiant companion. The same may be true for the He {\sc i} absorption feature near $\lambda$6680 \AA.}
\end{figure*}

\begin{figure*}[t]
\centering
\epsscale{1}
\plotone{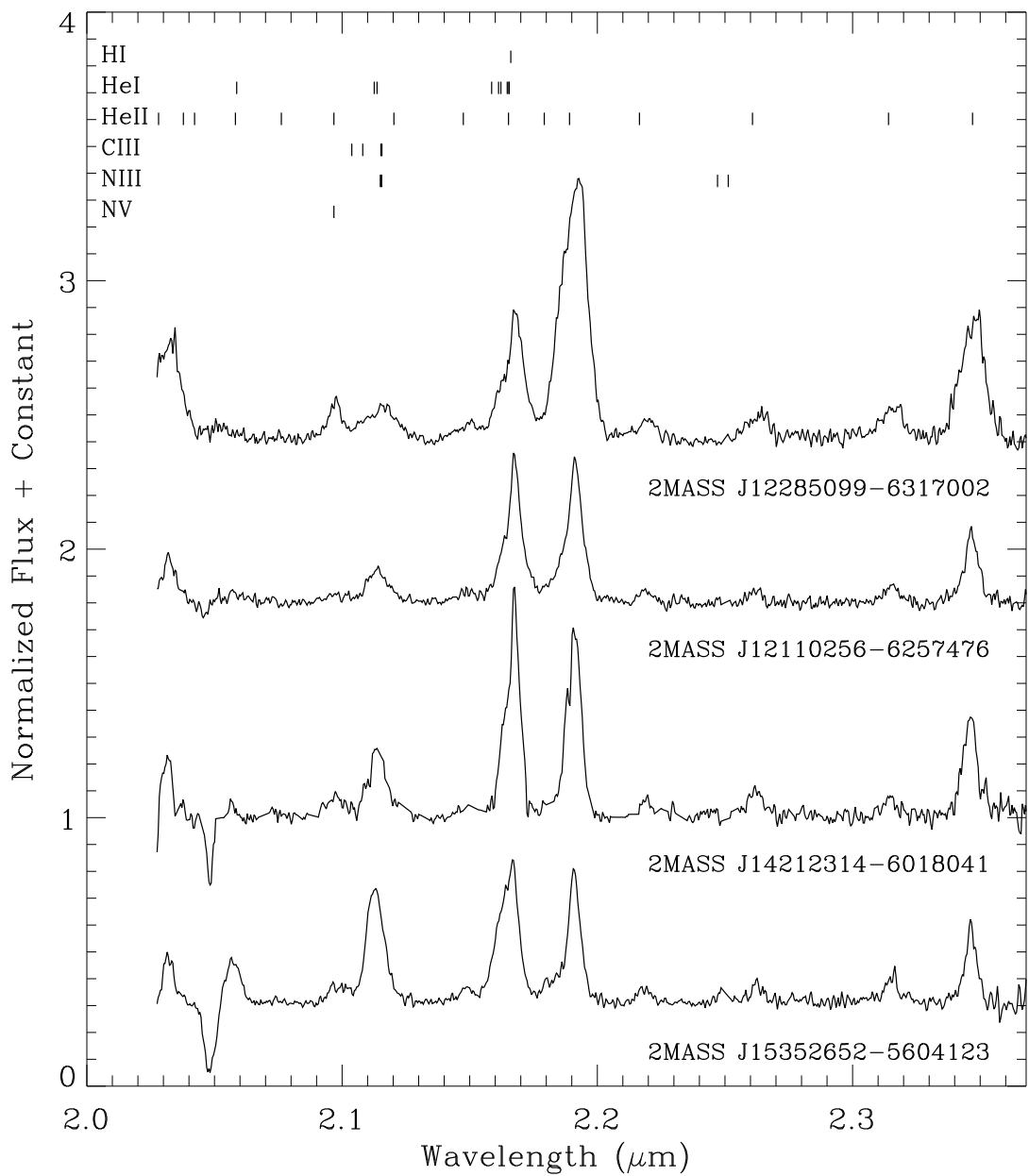}
\caption[]{\linespread{1}\normalsize{Near-infrared $K$-band spectra of new WRs  2MASS J12285099$-$6317002, 2MASS J12110256$-$6257476, 2MASS J14212314$-$6018041, and 2MASS J15352652$-$5604123 (upper to lower spectra, respectively), obtained with the AAT 3.9 m telescope and IRIS2 instrument. The relative strengths of He {\sc i}, He {\sc ii}, N {\sc iii} and N {\sc v} are consistent with nitrogen subtypes of WN4--5, WN6, WN6, and WN7 for the respective stars.}}
\label{archesquintuplet}
\end{figure*}

\begin{figure*}[t]
\centering
\epsscale{1}
\plotone{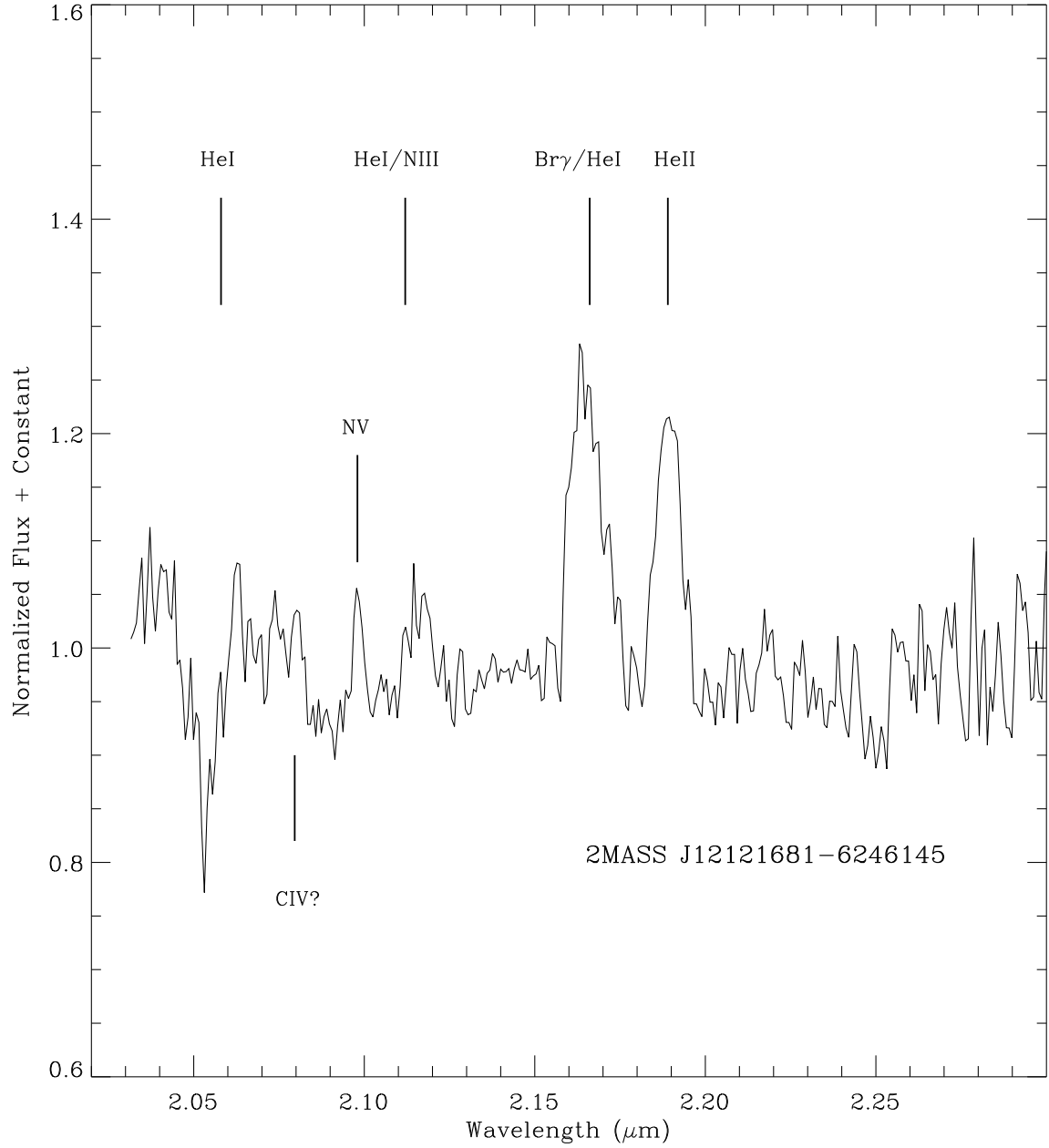}
\caption[]{\linespread{1}\normalsize{Near-infrared $K$-band spectrum of the WN4--6 star 2MASS J12121681$-$6246145, obtained with the SOAR 4.1 m telescope and OSIRIS instrument.}}
\end{figure*}

\begin{figure*}[t]
\centering
\epsscale{1}
\plotone{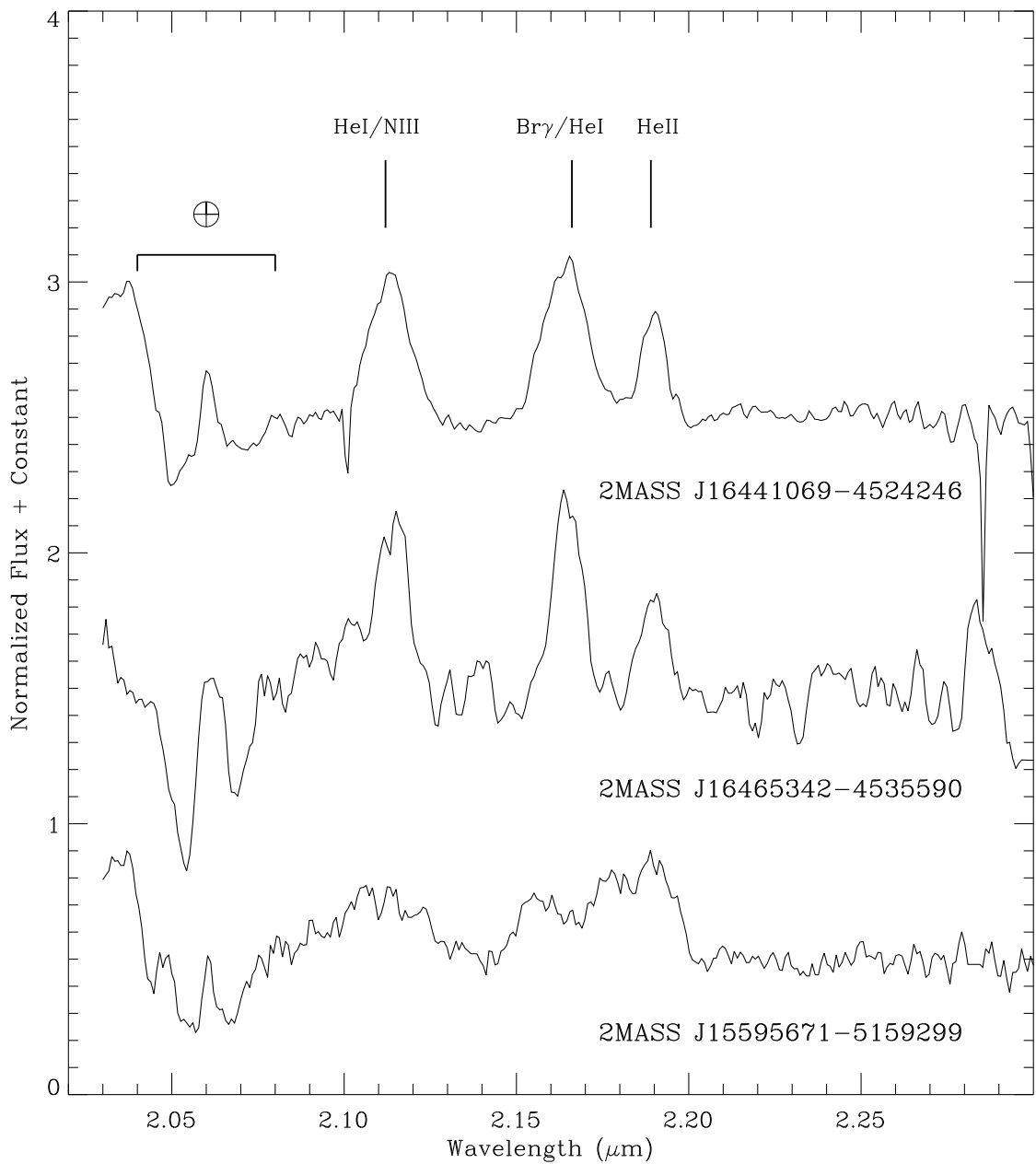}
\caption[]{\linespread{1}\normalsize{Near-infrared $K$-band spectra of stars 2MASS J16441069$-$4524246, 2MASS J16465342$-$4535590, and 2MASS J15595671$-$5159299, which we classify as WN7, WN7, and WN5--6b, respectively. These spectra were obtained with SOAR/OSIRIS in poor weather conditions, and lack telluric correction. Thus, we are unable to measure the strength of He {\sc i} ($\lambda$2.058 \micron), which is typically present in emission for WN5--7 stars.}}
\end{figure*}

\begin{figure*}[t]
\centering
\epsscale{1}
\plotone{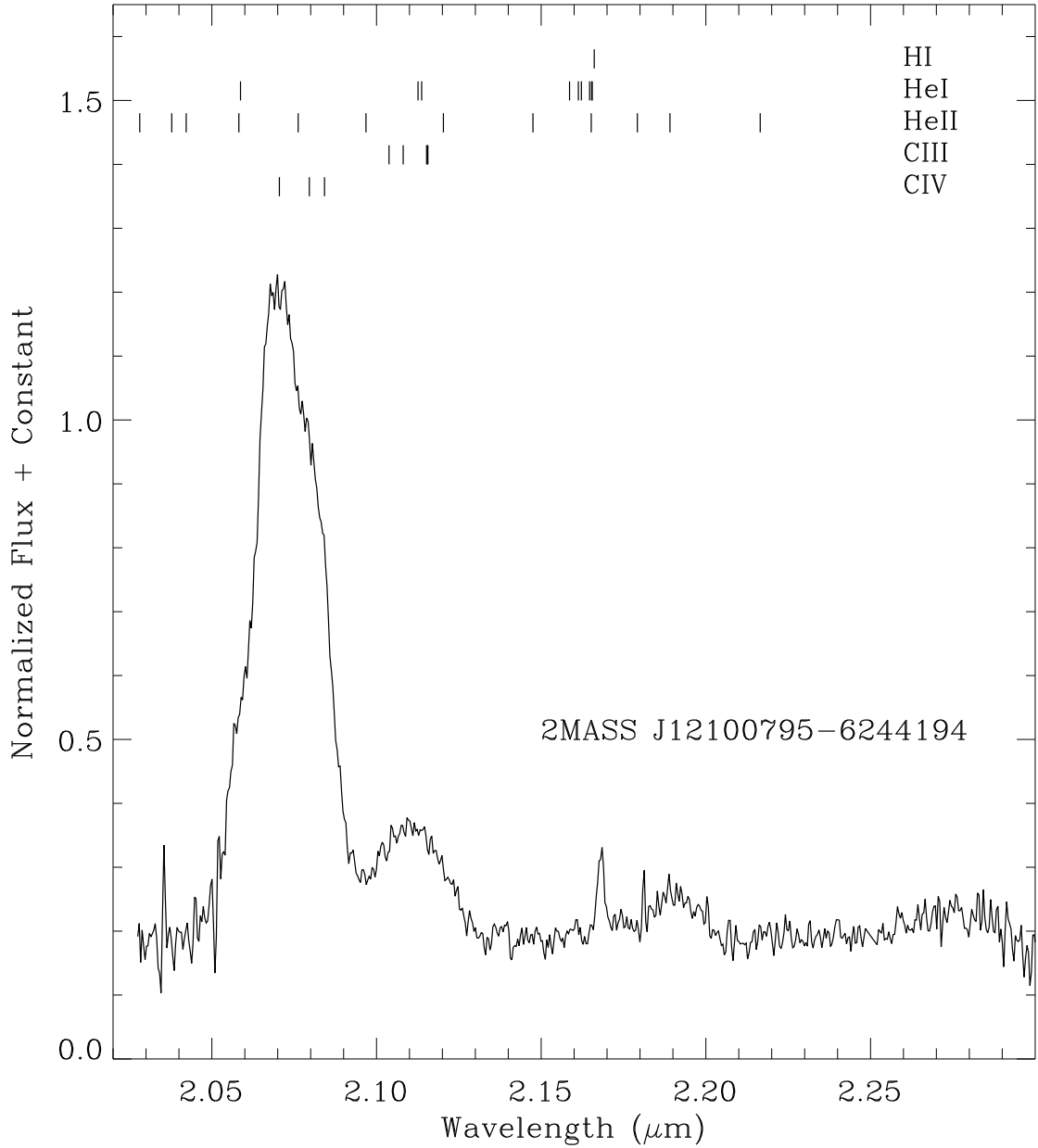}
\caption[]{\linespread{1}\normalsize{Near-infrared $K$-band spectrum of the WC5-7 star 2MASS J12100795$-$6244194, obtained with AAT/IRIS2. Very strong C {\sc iv} emission dominates this spectrum. Weaker lines of C {\sc iii} also contribute. He {\sc ii} is detected at $\lambda$2.189 {\micron}, in addition to relatively-narrow emission line of Br${\gamma}$, which may be nebular.}}
\end{figure*}

\begin{figure*}[t]
\centering
\epsscale{1}
\plotone{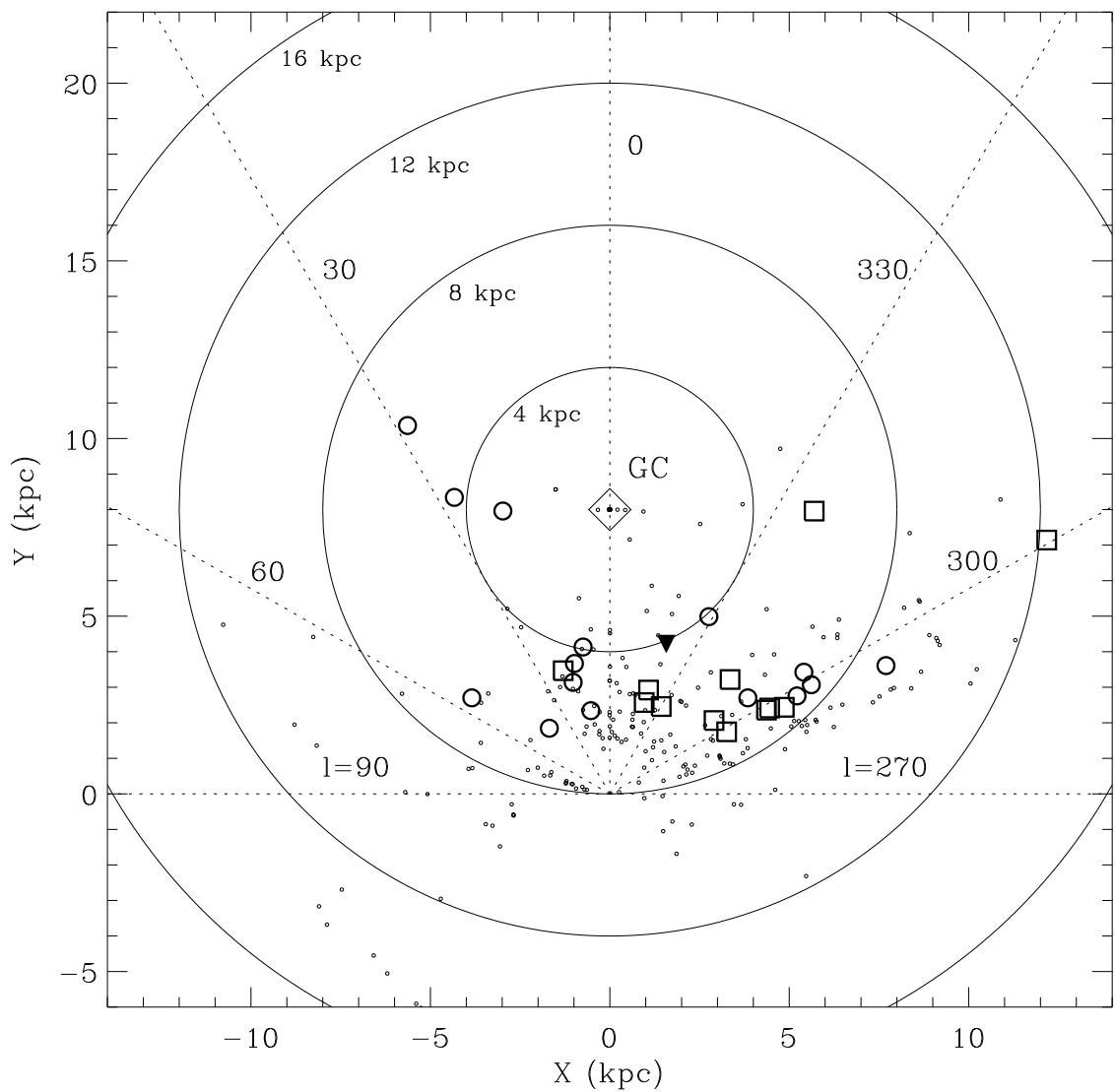}
\caption[]{\linespread{1}\normalsize{Galactic distribution of new WRs (\textit{squares}) and those from Hadfield et al. (2007, \textit{open circles}). The figure axes are in units of kpc, centered on the location of the Sun. The positions of known WRs from van der Hucht (2001, 2006) are marked (\textit{small dots}). The Galactic center (\textit{diamond symbol}), assumed to lie at 8.0 kpc, contains $\approx$20\% of the known Galactic WRs within 3 stellar clusters. The location of the WR-rich, starburst cluster Westerlund 1 (Crowther et al. 2006b) is also marked (\textit{inverted triangle}). Dotted lines mark several values of constant Galactic longitude. }}
\end{figure*}

\begin{figure*}[t]
\centering
\epsscale{1}
\plotone{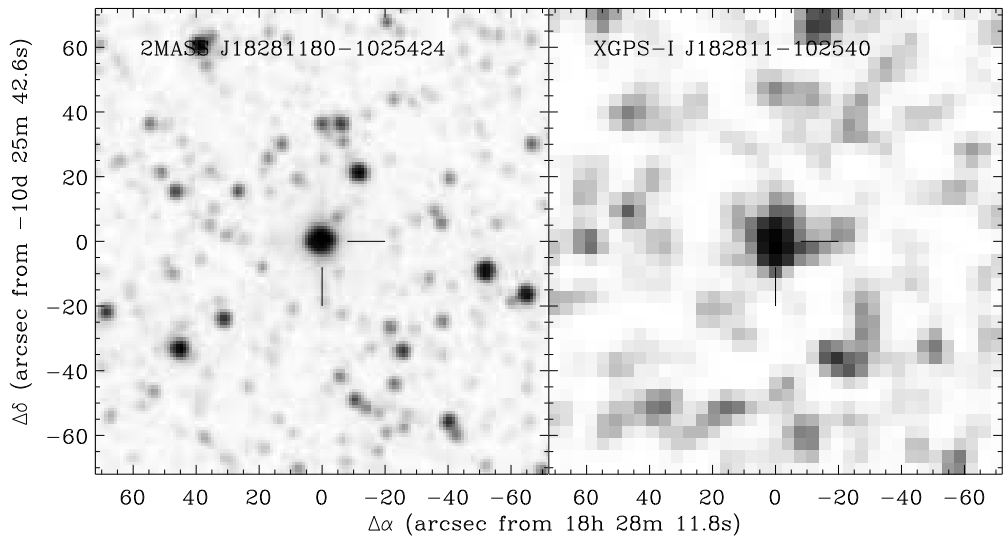}
\caption[]{\linespread{1}\normalsize{Images of the field containing the WC9d$+$OB{\sc I} binary 2MASS J18281180$-$1025424 (2MASS $K$-band in left panel), and the associated X-ray source XGPS-I J182811$-$102540 (\textit{XMM} 0.4--6.0 keV in right panel). Both images cover the same 144\arcsec~fields, and have north facing up and east toward the left. The \textit{XMM} image was obtained from the \textit{XMM-Newton} Science Archive and has been Gaussian-smoothed using a 2-pixel-radius filter.}}
\end{figure*}

\begin{figure*}[t]
\centering
\epsscale{1}
\plotone{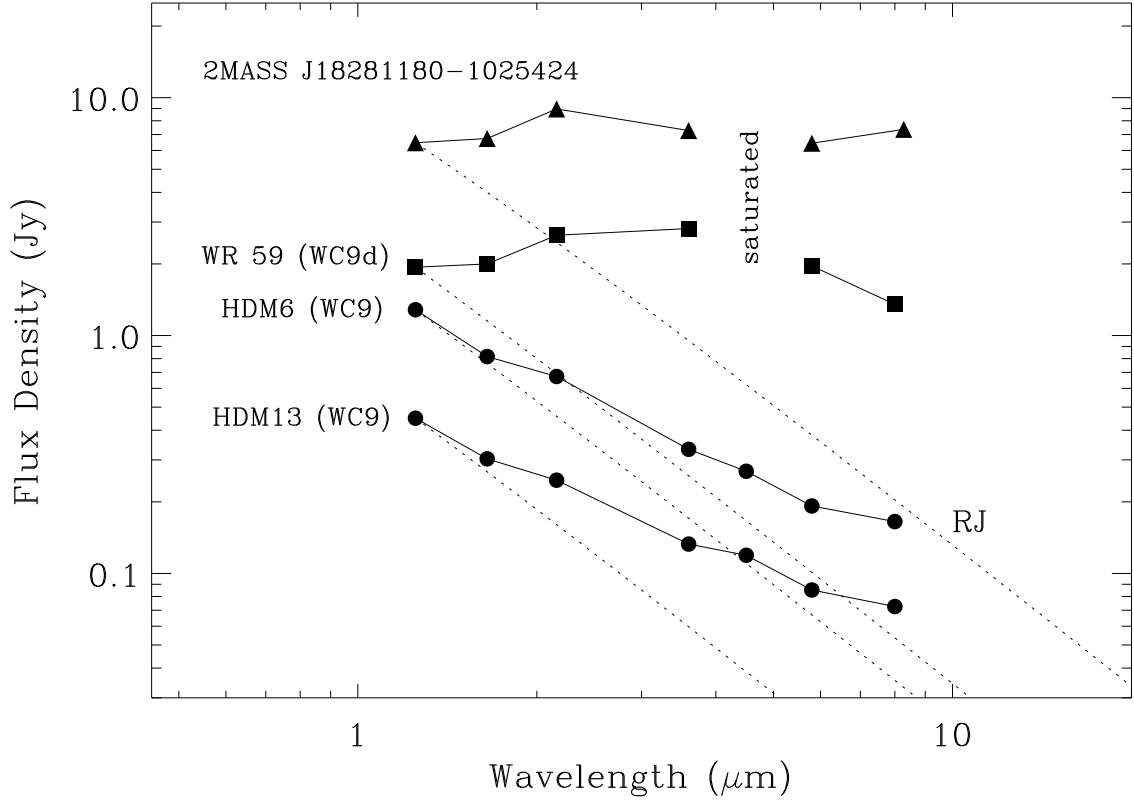}
\caption[]{\linespread{1}\normalsize{De-reddened spectral energy distributions (SED) of 2MASS J18281180$-$1025424 and several other known WC9 stars. Rayleigh-Jeans curves (\textit{dotted lines}) representing the stellar continuua of these stars are also plotted for comparison and are scaled to match the $J$-band flux. The infrared excess of J18281180$-$1025424 resembles that of the dusty WC9 star WR 59 (\textit{squares}), and so is likely to be comprised of both free-free and thermal dust emission. The SEDs of non-dusty WC9 stars HDM6 and HDM13 (\textit{circles}) from Hadfield et al. (2007) exhibit excess typical of free-free emission from WC9 winds, but lack an additional thermal dust components comparable to those of 2MASS J18281180$-$1025424 or WR 59.}}
\end{figure*}

\begin{figure*}[t]
\centering
\epsscale{1}
\plotone{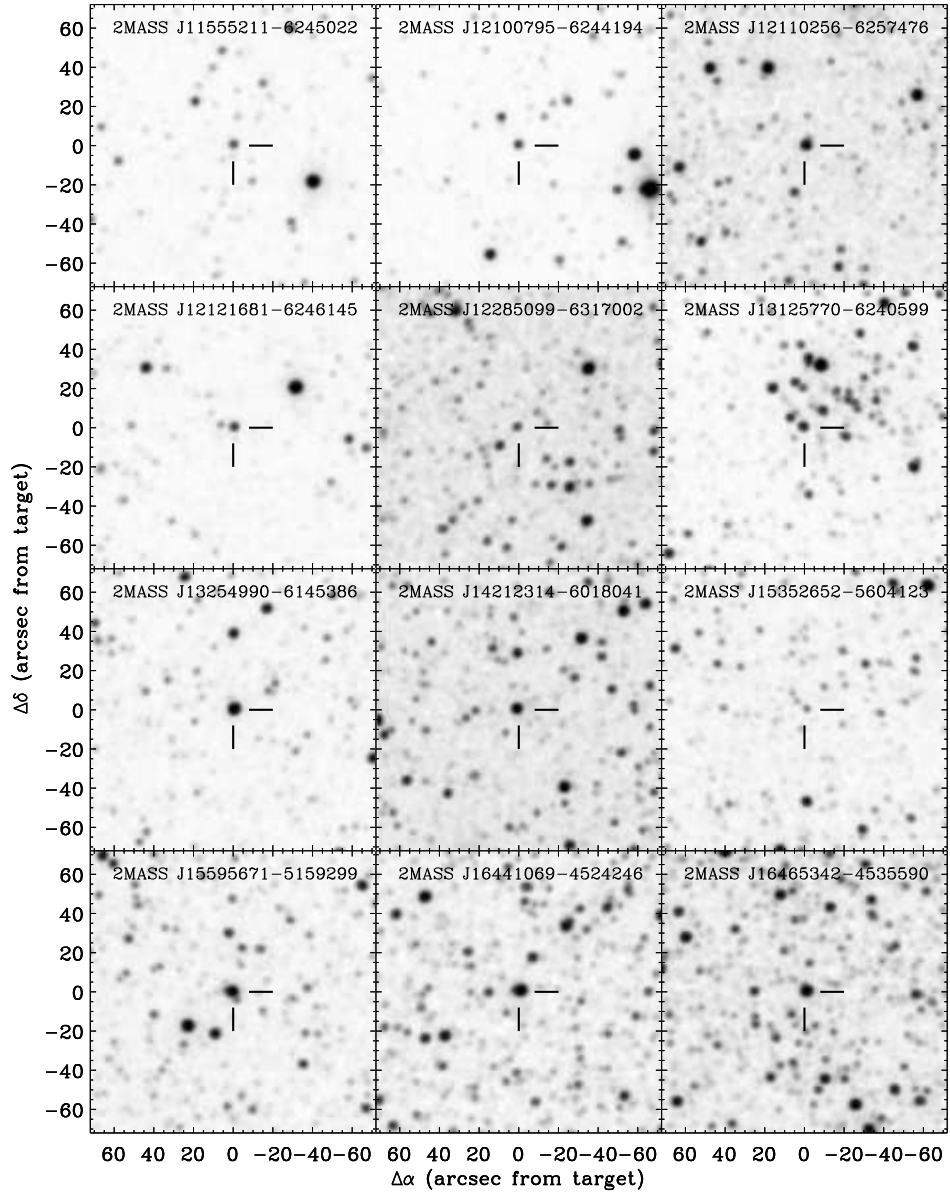}
\caption[]{\linespread{1}\normalsize{2MASS $K_s$-band images of the new WR fields. Each image is 144{\arcsec} on a side, and have north facing up and east toward the left.}}
\end{figure*}

\end{document}